\documentclass[12pt]{article}
\usepackage{fullpage}
\usepackage{amssymb, amsmath}
\usepackage{graphicx}
\usepackage{bbm}
\usepackage{appendix}

\usepackage[section]{placeins}
\makeatletter
\AtBeginDocument{%
\expandafter\renewcommand\expandafter\subsection\expandafter{%
\expandafter\@fb@secFB\subsection
}%
}
\makeatother

\newcommand{\argmin}{\operatorname*{arg \ min}}

\usepackage{algorithm,algorithmic}
\usepackage{setspace}

\usepackage{blindtext}
\usepackage[table]{xcolor}
\usepackage{xr}
\usepackage{booktabs}
\usepackage{breqn}
\usepackage[
backend=biber,
giveninits=true,
uniquename=false,
date=year,
style=authoryear-comp,
uniquelist=false,
sorting=nty,
url=false,
isbn=false,
doi=false,
eprint=false,
maxbibnames=1,
minbibnames=1,
maxcitenames=1,
mincitenames=1
]{biblatex}
\usepackage{tikz}
\usepackage{subfigure}
\usepackage{array}
\usepackage{enumerate}
\usepackage{amsthm}
\usepackage[shortlabels]{enumitem}
\newcolumntype{C}[1]{>{\centering\arraybackslash}p{#1}}
\usetikzlibrary{shapes.geometric}
\usetikzlibrary{arrows,automata}
\usepackage{soul}

\newtheorem{prop}{Proposition}

\newtheorem{lemma}{Lemma}
\DeclareMathOperator{\Y}{\mathbf{Y}}
\DeclareMathOperator{\G}{\mathbf{G}}
\DeclareMathOperator{\C}{\mathbf{C}}
\DeclareMathOperator{\E}{\mathbf{E}}
\DeclareMathOperator{\K}{\mathbf{K}}
\DeclareMathOperator{\Hn}{\mathbf{H}}
\DeclareMathOperator{\I}{\mathbf{I}}
\DeclareMathOperator{\bSigma}{\mathbf{\Sigma}}
\DeclareMathOperator{\vect}{vec}

\DeclareMathOperator{\diag}{diag}
\DeclareMathOperator{\Var}{Var}

\DeclareMathOperator{\cov}{cov}

\DeclareNameAlias{sortname}{last-first}

\title{Genetic Regression Analysis  of Human Brain Connectivity Using an Efficient Estimator of Genetic Covariance}

\author{Keshav Motwani, Ali Shojaie, Ariel Rokem, and Eardi Lila}

\date{}{}

\begin{document}
\def\spacingset#1{\renewcommand{\baselinestretch}%
{#1}\small\normalsize} \spacingset{1}

\maketitle

\vspace{-6mm}

\begin{abstract}

\noindent Non-invasive measurements of the human brain using magnetic resonance imaging (MRI) have significantly improved our understanding the brain's network organization by enabling measurement of anatomical connections between brain regions (structural connectivity) and their coactivation (functional connectivity).
Heritability analyses have established that genetics account for considerable intersubject variability in structural and functional connectivity. However, characterizing how genetics shape the relationship between structural and functional connectomes remains challenging, since this association is obscured by unique environmental exposures in observed data. To address this, we develop a regression analysis framework that enables characterization of the relationship between latent genetic contributions to structural and functional connectivity. Implementing the proposed framework requires estimating genetic covariance matrices in multivariate random effects models, which is computationally intractable for high-dimensional connectome data using existing methods. We introduce a constrained method-of-moments estimator that is several orders of magnitude faster than existing methods without sacrificing estimation accuracy. For the genetic regression analysis, we develop regularized estimation approaches, including ridge, lasso, and tensor regression. Applying our method to Human Connectome Project data, we find that functional connectivity is moderately predictable from structure at the genetic level (max $R^2
= 0.34$), though it is not directly predictable in the observed data (max $R^2 = 0.03$). This stark contrast suggests that unique environmental factors mask strong genetically-encoded structure-function relationships.
\end{abstract}



\spacingset{1.65} 

\vspace{-3.5mm}

\section{Introduction}

Striking levels of intersubject heterogeneity have been observed in the structure and function of the human brain, even within narrowly defined groups of young and healthy subjects \parencite{benson2022variability}. Brain connectivity is no exception, suggesting that a large variety of connectivity configurations can result in cognitively healthy subjects \parencite{finn2015functional}. Two types of connectivity are commonly studied: \emph{structural connectivity}, which describes anatomical connections between brain regions through white matter fibers, and \emph{functional connectivity}, which measures the degree of co-activation of brain regions. The set of connections between pairs of macroscopic brain regions is referred to as the connectome.

Structural connectivity is believed to provide the physical substrate supporting functional organization. This structure–function coupling has been extensively characterized \parencite{baum2020development, pan2025mapping, fotiadis2024structure}. However, while it is well established that both structural and functional connectivity are influenced by genetic \parencite{colclough2017heritability,elliott2018genomewide,gu2021heritability,bianco2023heritability} and environmental factors \parencite{kulick2020longterm,pini2023pollutome}, the specific aspects of structure–function coupling that are genetically rooted versus those shaped by environmental exposures are not yet well understood. Classical analyses of structure–function coupling in neuroimaging studies may mask underlying relationships when genetic and environmental factors contribute differently to this coupling. On the other hand, univariate heritability analyses are inadequate for addressing this question because they fail to capture the complex and multivariate nature of structure–function coupling.

In the present work, we developed a \textit{latent genetic regression analysis} framework that characterizes genetically encoded structure-function associations. Specifically, we represent observed connectomes as the sum of latent genetic and environmental components and model the genetic component of the functional connectome as a linear function of the genetic component of the structural connectome. To estimate this linear mapping, we develop a moment-matching genetic covariance estimator with a positive semidefinite constraint, which we use as a plug-in to estimate the linear regression coefficients within a ridge, lasso, or tensor regression framework. Our analysis reveals that functional connectomes show substantially higher predictability from structural connectomes at the genetic component level than at the phenotypic level, suggesting that strong genetically-encoded structure-function relationships exist but are masked by environmental heterogeneity. 

\begin{figure}[!t]
\centering
\includegraphics[width=\textwidth]{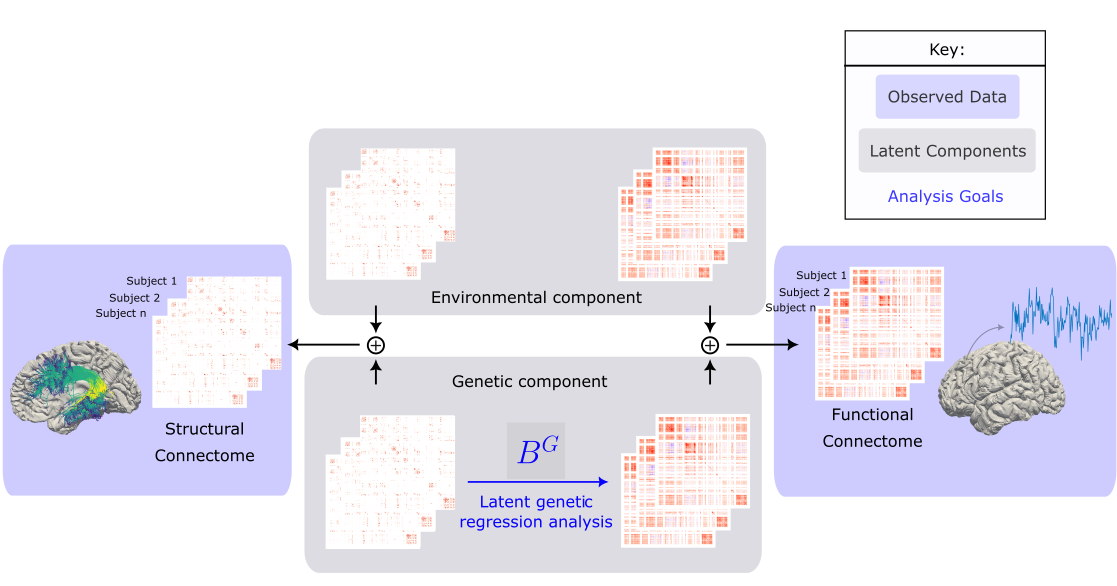}
\caption{Illustration of the proposed approach. Observed connectomes are modeled as the sum of latent genetic and environmental components. Our methodology characterizes the relationship between structural and functional connectivity within the genetic component through the proposed latent genetic regression analysis framework.}\label{fig:introfig}
\end{figure}

\vspace{-3.5mm}

\subsection{Motivating data and statistical analysis}\label{sec:data}

In this work, we use data from the Human Connectome Project (HCP), which contains multimodal imaging data from 1,200 young adults, including diffusion-weighted MRI for structural connectivity and resting-state functional MRI \parencite{vanessen2012human}. Critically, the dataset includes genetically verified family structures (149 monozygotic twin pairs, 94 dizygotic twin pairs), enabling identification of genetic and environmental contributions.


Figure~\ref{fig:introfig} illustrates our analysis. Structural connectomes are constructed from diffusion MRI using tractography to count streamlines connecting each pair of regions. Functional connectomes are constructed from temporal correlations of cortical activation between regions. As we describe in Section~\ref{subsec:datanalaysisresults}, when we perform classical linear regression to predict functional connectivity from structural connectivity in the observed data, we find weak associations (maximum observed $R^2$ = 0.03). This limited predictability motivates our investigation of whether stronger structure-function associations exist at the genetic level but are masked by environmental effects. We therefore model the genetic component of functional connectivity as a linear function of the genetic component of structural connectivity, with regression coefficients capturing intrinsic structure-function associations independent of environmental factors.

Implementing the proposed genetic regression analysis at connectome scale presents a major computational challenge. The analysis requires estimating the joint genetic covariance matrix of structural and functional connectomes, from which the regression parameters can be estimated. Under our modeling choices, the structural and functional connectomes together constitute a 4,624-dimensional phenotype, which existing likelihood-based methods cannot handle. We therefore developed a computationally efficient covariance estimator that renders the proposed analysis tractable and enables the use of high-dimensional regression frameworks for our genetic regression analysis.

\vspace{-3.5mm}

\subsection{Related literature}\label{sec:litreview}





Our proposed approach bridges methods for heritability analysis and estimation of variance components. Therefore, in this section we review relevant literature from these fields.
Heritability analyses assess the extent to which trait variation is attributable to genetic factors, typically using additive genetics and common environment (ACE) models \parencite{maes2005ace}. These models are an application of linear mixed models, which partition the total phenotypic variance by including random effects for additive genetic (A), common environment (C), and unique environment (E) components, each defined by its expected covariance structure. Restricted maximum likelihood (REML) is the gold standard for variance component estimation, but it is computationally prohibitive for multivariate traits. Method-of-moments estimators are faster \parencite{lindquist2012estimating} but may produce non-positive estimates. Recently, constrained method-of-moments-based estimators that guarantee nonnegative variance estimates have been proposed and studied in \textcite{zou2017covariance} and \textcite{yue2021rehe}; these methods improve upon estimation performance of their unconstrained variants while matching their computational efficiency.

Previous work on structure-function coupling \parencite{gu2021heritability} investigated the role of genetics by first estimating subject-specific structure-function correlations and then assessing the heritability of this univariate trait. This approach enables the use of univariate heritability models, however, it does not directly model population-level relationships between genetic components of structural and functional connectivity.

Genetic contributions to multivariate traits, such as entire connectomes, have been typically investigated using massive univariate analyses, in which heritability is estimated separately for each trait. For example, numerous heritability studies of brain connectivity have analyzed one brain connection at a time using a univariate model; see, e.g., \textcite{colclough2017heritability,ge2017heritability} and \textcite{elliott2018genomewide}. 
Extensions to multidimensional imaging phenotypes have recently been proposed \parencite{tian2023bayesian, zhao2022bayesian, zhao2022genetic}. While these approaches improve estimation accuracy by borrowing information from neighboring imaging regions, they mostly focus on estimating heritability and not the genetic and environmental cross-covariance between distinct regions. Other multidimensional extensions are specifically designed for twin study designs \parencite{risk2021ace, luo2019fsem}. These methods either rely on a truncation approach to constrain the covariance estimate to be positive semi-definite, which may be suboptimal, or require estimating the covariance rank from the scree plot, a non-trivial task.

Massively univariate analyses do not estimate the covariance matrices of the genetic component, and thus do not provide any biological insight into the multidimensional structure of the genetic contributions. Since several multivariate statistical estimands are functions of the population covariance matrix,  estimating the entire \textit{covariance matrix} of the genetic component enables a large class of downstream analyses on the latent genetic component even though it cannot be directly observed. Although we focus here on linear regression to characterize the genetic structure-function coupling, our framework also facilitates other methods, such as principal component analysis and canonical correlation analysis.


Unlike their univariate counterparts, the estimation of the covariance matrix of latent multi-dimensional components has received limited attention. REML estimation of the covariance matrices using \texttt{lme4} \parencite{bates2013fast} is computationally prohibitive with as few as 10 traits and 1,000 subjects. \textcite{zhou2014efficient} proposed more efficient algorithms for computing the REML estimator, though their computational approach does not extend to models with more than two latent components, such as the additive genetics and common environment model considered in this work. Commonly used software packages, such as GCTA \parencite{yang2011gcta}, only support up to two traits. To get around the computational bottleneck associated with ML-based estimators, method-of-moments-based estimators have been proposed \parencite{ge2016multidimensional, liegeois2019resting, anderson2021heritability}. However, these estimators have poor estimation performance compared to REML, particularly when using familial data in the additive genetics and common environment model. This is due to the high similarity between the kinship and shared environment indicator matrices, which is especially relevant in the HCP data analyzed in this work \parencite{ge2016multidimensional}. Furthermore, these methods may produce covariance estimates that are not positive semi-definite, complicating downstream analyses of the latent components. 
\vspace{-3.5mm}

\subsection{Our contributions}

We introduce a latent genetic regression analysis framework that characterizes genetically-encoded structure-function associations by modeling relationships between genetic components of structural and functional connectivity. Implementing this analysis at the connectome scale requires efficient estimation of genetic and environmental covariance matrices. To this end, we introduce a constrained method-of-moments estimator that guarantees positive semidefinite estimates, closely matches REML’s accuracy, and provides substantial computational speedups. For $q$ connections and $n$ samples, our method requires $\mathcal{O}(n^2 q^2)$ operations upfront followed by $\mathcal{O}(\min(n, q)^3)$ operations per iteration, compared to at least $\mathcal{O}(n^3 q^3)$ operations per iteration for REML \parencite{bates2015fitting}.

To implement the proposed genetic regression analysis, we developed regularized estimation approaches based on plug-in estimates of the genetic covariance matrix, including ridge regression, lasso regression, and tensor regression that leverages connectome structure. These methods estimate linear mappings from the latent genetic component of the structural connectome to the latent genetic component of the functional connectome, without requiring direct observation of the genetic component itself.

Applied to HCP data, we find functional connectivity is substantially more predictable from structural connectivity at the genetic level than in observed data (maximum $R^2$: 0.34 vs 0.02). This stark contrast suggests unique environmental factors mask genetically-encoded relationships, and the uncovered genetically-encoded relationships may provide insights into the biological architecture of structure-function coupling.

The remainder of the paper is organized as follows. Section~\ref{sec:methods} formulates the genetic regression framework and introduces our proposed genetic covariance estimator. Section~\ref{sec:application} applies our methodology to HCP data. Section~\ref{sec:simulations} presents extensive simulations validating our approach. We end with some concluding remarks in Section~\ref{sec:discussion}. Supplementary Materials contain the derivation of the numerical algorithm, asymptotic theory, additional simulations, and extensions to function-valued phenotypes.

\vspace{-3.5mm}

\section{Methodology}\label{sec:methods}
\vspace{-3.5mm}

\subsection{Latent genetic regression analysis}


To disentangle genetic and environmental factors in shaping brain connectivity, we adopt a multivariate additive genetics and common environment model, in which a  $q$-dimensional trait $\mathbf{y}_i$ is modeled as the sum of independent genetic, common environment, and unique environment random components:
\begin{equation}  \label{eq:multivariateindivmodel2}
\mathbf{y}_i = \mathbf{g}_i + \mathbf{c}_i + \mathbf{e}_i
\end{equation}
where
$\mathbf{g}_i \sim \mathcal{N}(0, \bSigma_G)$, $\mathbf{c}_i \sim \mathcal{N}(0, \bSigma_C)$, and $\mathbf{e}_i \sim \mathcal{N}(0, \bSigma_E)$. Here, $\mathcal{N}(0, \bSigma)$ denotes the multivariate normal distribution with mean $0$ and covariance matrix $\bSigma$ and $\bSigma_G, \bSigma_C$ and $\bSigma_E$ are unknown $q \times q$ covariance matrices. Specifically, we model the vectorized structural and functional connectomes as a multivariate trait $\mathbf{y}_i = \left(\mathbf{y}_i^{\mathcal{S}}, \mathbf{y}_i^{\mathcal{F}} \right)$ and partition the latent components conformably, e.g., $\mathbf{g}_i = \left(\mathbf{g}_i^{\mathcal{S}}, \mathbf{g}_i^{\mathcal{F}} \right)$.

To characterize the association between structural and functional brain connections driven by genetic factors, our primary goal is to perform linear regression of the functional connectome on the structural connectome, restricted to the genetic component of each. Specifically, we fit separate regressions for each functional connection, using the full structural connectome as the predictor. Formally, for the $j$th functional connection, we are interested in the best linear approximation of its genetic component based on the genetic component of the entire structural connectome:
\begin{equation} \label{eq:geneticbeta}
\boldsymbol{\beta}_{G, j}^* = \argmin_{\boldsymbol{\beta}} \mathbb{E}\left[ \left\|\mathbf{g}_{i}^{\mathcal{F}_j} - \boldsymbol{\beta}^\top \mathbf{g}_{i}^{\mathcal{S}} \right\|_2^2\right]
\end{equation}
and the associated proportion of genetic variance explained (genetic-$R^2$):
\begin{equation} \label{eq:geneticr2}
R^2_{G, j}(\boldsymbol{\beta}) = 1 - \frac{\Var\left(\mathbf{g}_{i}^{\mathcal{F}_j} - \boldsymbol{\beta}^\top \mathbf{g}_{i}^{\mathcal{S}}\right)}{\Var\left(\mathbf{g}_{i}^{\mathcal{F}_j}\right)}.
\end{equation}

These genetic regression coefficients capture the intrinsic associations between structural and functional connectomes that exist independent of environmental influences, representing the genetic ``blueprint'' of structure-function relationships.

In previous studies of structure–function coupling \parencite{gu2021heritability}, the coupling is typically characterized by a subject-specific parameter—such as the correlation between the observed structural and functional connectomes—and the heritability of these parameters is reported. These approaches implicitly assume that each functional connection influences only its corresponding structural connection to enable the estimation of subject-specific correlations. In contrast, genetic $R^2$ captures more complex population-level relationships between structure and function. Additional details on the interpretation of these parameters in the context of structure–function coupling studies are left to Section~\ref{sec:application}.

Although the genetic regression and genetic $R^2$ parameters are defined in terms of latent unobservable variables, our proposed analysis leverages the observation that they can be expressed in terms of specific blocks of the genetic covariance matrix, $\bSigma_G$. We partition the genetic covariance matrix according to the structural and functional components:
$$
\bSigma_G = 
\begin{bmatrix}
\bSigma_G^{\mathcal{S},\mathcal{S}} & \bSigma_G^{\mathcal{S},\mathcal{F}} \\
\bSigma_G^{\mathcal{F},\mathcal{S}} & \bSigma_G^{\mathcal{F},\mathcal{F}}
\end{bmatrix}. 
$$
The genetic regression parameter can then be identified as
\begin{equation}\label{eq:geneticbeta_cov}
\boldsymbol{\beta}_{G, j}^*  = 
\argmin_{\boldsymbol{\beta}} \boldsymbol{\beta}^\top \bSigma_G^{\mathcal{S}, \mathcal{S}} \boldsymbol{\beta} - 2\boldsymbol{\beta}^\top \bSigma_G^{\mathcal{S},\mathcal{F}_j},
\end{equation}
and the genetic $R^2$ as
\begin{equation}\label{eq:geneticr2_cov}
R^2_{G, j}(\boldsymbol{\beta}) = 1 - \frac{\bSigma_G^{\mathcal{F}_j,\mathcal{F}_j} - 2 \boldsymbol \beta^\top \bSigma_G^{\mathcal{S}, \mathcal{F}_j} + \boldsymbol \beta^\top \bSigma_G^{\mathcal{S}, \mathcal{S}} \boldsymbol \beta }{\bSigma_G^{\mathcal{F}_j, \mathcal{F}_j}}.
\end{equation}

Therefore, estimating the genetic covariance matrix, $\bSigma_G$, enables plug-in estimation of the genetic regression parameters. The plug-in loss minimization framework naturally accommodates regularized approaches, including ridge and lasso regression, as well as extensions to tensor regression that respect the matrix structure of the structural connectome predictor. We first introduce our efficient covariance estimation methodology in the next section, then develop the latent regression estimation methodology in Section~\ref{sec:regression_estimation}.

\subsection{Estimation of covariance matrices}
\subsubsection{Observed data model}
Let $\mathbf{y}_1, \dots, \mathbf{y}_n$ be $n$ identically distributed observations according to the model in \eqref{eq:multivariateindivmodel2} and define $\Y \in \mathbb{R}^{n \times q}$ as the matrix with rows $\mathbf{y}_i^\top$. Similarly defining matrices $\G$, $\C$, and $\E$ for the genetic, common environmental, and unique environmental components across all subjects, we can write
\begin{equation} \label{eq:multivariatevariateobservedmodel}
\Y = \G + \C + \E.
\end{equation}
For an $n \times q$ matrix-valued random variable $\mathbf{Z}$, we say $\mathbf{Z} \sim \mathcal{MN}(\mathbf{0}, \mathbf{D}, \bSigma)$ follows a matrix-normal distribution with row-covariance $\mathbf{D}$ (describing dependence between subjects) and column-covariance $\bSigma$ (describing dependence between traits). This is formally defined as $\vect(\mathbf{Z}) \sim \mathcal{N}(0, \bSigma \otimes \mathbf{D})$, where $\vect(\cdot)$ denotes vectorization and $\otimes$ is the Kronecker product.

If observations were independent across subjects, we would have $\G \sim \mathcal{MN}(\mathbf{0}, \mathbf{I}_n, \bSigma_G)$, $\C \sim \mathcal{MN}(\mathbf{0}, \mathbf{I}_n, \bSigma_C)$, and $\E \sim \mathcal{MN}(\mathbf{0}, \mathbf{I}_n, \bSigma_E)$. However, since the row-covariances are the same, only the sum $\bSigma_G + \bSigma_C + \bSigma_E$ is identifiable from the data---the individual covariance matrices cannot be separately estimated.

Identifying the individual covariance matrices therefore requires an experimental design that includes related subjects with varying degrees of genetic relatedness and shared environmental exposure. Let $\K_n$ and $\Hn_n$ be $n \times n$ positive semi-definite matrices with diagonal entries equal to 1, where off-diagonal entries quantify the genetic relatedness and shared environment status between each pair of subjects, respectively:
\begin{equation}
\label{eq:multivariatevariateobservedmodel2}
\G \sim \mathcal{MN}(\mathbf{0}_{n \times q}, \K_n, \bSigma_G), \quad \C \sim \mathcal{MN}(\mathbf{0}_{n \times q}, \Hn_n, \bSigma_C), \quad \E \sim \mathcal{MN}(\mathbf{0}_{n \times q}, \I_n, \bSigma_E).
\end{equation}

Under the model in \eqref{eq:multivariatevariateobservedmodel} and \eqref{eq:multivariatevariateobservedmodel2}, the individual covariance matrices $\bSigma_G$, $\bSigma_C$, and $\bSigma_E$ become identifiable provided that $\K_n$, $\Hn_n$, and $\mathbf{I}_n$ are linearly independent, which is satisfied when the study includes subjects with varying degrees of genetic relatedness and shared environmental exposure.

Existing methods for estimation of the covariance matrices in the multivariate random effects model are computationally expensive, especially with a large number of traits. To overcome this challenge, we propose an extension of \textcite{yue2021rehe} to the multivariate setting, using a constrained method-of-moments estimator rather than maximum likelihood. Our approach requires only the first two moments of the data, specifically the Kronecker covariance structure: e.g. $\cov(\vect(\G))=\bSigma_G \otimes \K_n$. Therefore, while normal distributions are used for expositional clarity, the normality assumption is not required for our estimation procedure.

\vspace{-3.5mm}

\subsubsection{Restricted multivariate Haseman-Elston regression}

The model introduced is a special case of the more general model with $K + 1$ components:
\begin{equation} \label{eq:genmodel}
\Y = \sum_{k = 0}^K \boldsymbol{\Gamma}_k,
\end{equation}
where $\boldsymbol{\Gamma}_k \sim \mathcal{MN}(\mathbf{0}_{n \times q}, \mathbf{D}_k, \bSigma_k)$. The model in (\ref{eq:multivariatevariateobservedmodel}) can be recovered by setting $\mathbf{D}_0 = \I_n$, $\mathbf{D}_1 = \K_n$ and $\mathbf{D}_2 = \Hn_n$ so that $\G = \boldsymbol{\Gamma_1}$, $\C = \boldsymbol{\Gamma_2}$, and $\E = \boldsymbol{\Gamma_0}$.

Classical Haseman-Elston (HE) regression \parencite{haseman1972investigation} estimates the variance components in a univariate model using a generalized method-of-moments estimator that minimizes the squared difference between theoretical and empirical second moments. To generalize this to the multivariate setting, we use the fact that the second moments of $\Y$ under the model in (\ref{eq:genmodel}) are given by
\begin{equation} \label{eq:ee}
\mathbb{E}[\Y_{i, j} \Y_{l, m}] = \sum_{k = 0}^K [\bSigma_k]_{j, m} [\mathbf{D}_k]_{i, l},
\end{equation}
for all $i,l \in [n], j,m \in [q]$. 
Therefore, a multivariate extension of HE regression would minimize
\begin{align} 
\hat \bSigma_0, \dots, \hat \bSigma_K &= \argmin_{\bSigma_k} \sum_{j = 1}^q \sum_{m = 1}^{q} \sum_{i = 1}^n \sum_{l = 1}^{n} \left(\Y_{i,j}\Y_{l,m} - \sum_{k = 0}^K [\bSigma_k]_{j, m} [\mathbf{D}_k]_{i, l} \right)^2 \label{eq:mvHE} \\
&= \argmin_{\bSigma_k} \sum_{j = 1}^q \sum_{m = 1}^{q} \left\|\Y_{:,j} \Y_{:,m}^\top   -  \sum_{k = 0}^K [\bSigma_k]_{j, m} \mathbf{D}_k \right\|_F^2. \label{eq:leastsquares}
\end{align}
The optimization problem \eqref{eq:leastsquares} can be solved for each pair of outcome variables $(j, m) \in [q] \times [q]$ separately, which can be rewritten as a least-squares linear regression. Specifically, let
$
\tilde{\Y}^{(j, m)} = \vect(\Y_{:, j} \Y_{:, m}^\top),
$
$
\tilde{\mathbf{X}} = (\vect(\mathbf{D}_0), \dots, \vect(\mathbf{D}_K)),
$
and
$
{\sigma}^{(j, m)} = ([\bSigma_0]_{j, m}, \dots, [\bSigma_K]_{j, m})^\top
$. Then solving (\ref{eq:leastsquares}) is equivalent to solving
\begin{equation*} \label{eq:min}
\hat \sigma^{(j, m)} = \argmin_{\sigma^{(j, m)}} \left\| \tilde{\Y}^{(j, m)} - \tilde{\mathbf{X}} {\sigma}^{(j, m)}\right\|_2^2
\end{equation*} with $[\hat{\bSigma}_k]_{j, m} = [\hat{\sigma}^{(j, m)}]_{k+1}$. This problem has a closed-form solution given by
$
\hat{\sigma}^{(j, m)} = (\tilde{\mathbf{X}}^{\top} \tilde{\mathbf{X}})^{-1} \tilde{\mathbf{X}}^{\top} \tilde{\Y}^{(j, m)}.
$

The estimator in \eqref{eq:leastsquares} has been proposed before by \textcite{ge2016multidimensional} for heritability analyses of neuroimaging traits and has been used by \textcite{liegeois2019resting} and \textcite{anderson2021heritability}. However, this estimator does not return positive semi-definite (PSD) estimates in general, requiring ad-hoc corrections before performing analyses using the covariance matrix. A simple correction, analogous to truncating the variance estimates at $0$ in the univariate case, is to truncate the eigenvalues of $\hat{\bSigma}_k$ at $0$. We refer to the truncated version of \eqref{eq:leastsquares} as the multivariate Haseman-Elston (mvHE) estimator.

Although truncating the estimate obtained from \eqref{eq:leastsquares} results in positive semi-definite estimates of the covariance matrices, these estimates do not minimize \eqref{eq:leastsquares} subject to the constraint that each of the $\bSigma_k$ is positive semi-definite (PSD). Thus, we consider solving \eqref{eq:leastsquares} with positive semi-definiteness constraints, which we call the restricted multivariate Haseman-Elston (mvREHE) estimator:
\begin{align}
\hat \bSigma_0, \dots, \hat \bSigma_K &= \argmin_{\bSigma_k \succeq 0} \sum_{j = 1}^q \sum_{m = 1}^{q} \left\|\Y_{:,j} \Y_{:,m}^\top   -  \sum_{k = 0}^K [\bSigma_k]_{j, m} \mathbf{D}_k \right\|_F^2 \label{eq:mvREHEunpen1} \\
&= \argmin_{\bSigma_k \succeq 0}  \sum_{i = 1}^n \sum_{l = 1}^{n} \left\|\Y_{i,:}^\top \Y_{l,:}   - \sum_{k = 0}^K [\mathbf{D}_k]_{i, l} \bSigma_k \right\|_F^2. \label{eq:mvREHEunpen}
\end{align}

The optimization problem in \eqref{eq:mvREHEunpen1} is no longer separable across pairs of outcome variables due to the PSD constraint. Therefore, to solve this optimization problem, we use a block coordinate descent algorithm. Specifically, we update $\bSigma_z$ with $\bSigma_k$ for $k \neq z$ held fixed at $\hat \bSigma_k^{(t)}$:
\begin{equation} \label{eq:update}
\hat \bSigma_z^{(t+1)} = \argmin_{\bSigma_z \succeq 0} \sum_{i = 1}^n \sum_{l = 1}^{n} \left\|\left(\Y_{i,:}^\top \Y_{l,:} - \sum_{k\neq z} [\mathbf{D}_k]_{i, l} \hat \bSigma_k^{(t)} \right) - [\mathbf D_k]_{i, l} \bSigma_z \right\|_F^2.
\end{equation}
Since the objective function in (\ref{eq:mvREHEunpen}) is convex and differentiable, with each of the blocks constrained to a convex set, block coordinate descent updates will converge to the solution of (\ref{eq:mvREHEunpen}).
The following lemma, proved in Appendix \ref{prooflemma}, allows us to compute each of these updates in closed form.
\begin{lemma} \label{lemma:update}
The solution to 
$$
\hat \bSigma = \argmin_{\bSigma \succeq 0}  \sum_{w = 1}^W \left\|\mathbf{S}_w - a_w \bSigma \right\|_F^2
$$
is $\hat \bSigma = \mathbf U \mathbf \Lambda^+ \mathbf U^\top$, where $\mathbf U \mathbf \Lambda \mathbf U^\top$ is the eigendecomposition of 
$$
\mathbf{S} = \left( \sum_{w = 1}^W a_w^2 \right)^{-1} \sum_{w = 1}^W a_w \mathbf{S}_w
$$
and $\mathbf \Lambda^+_{j,j} = \max(\mathbf \Lambda_{j,j}, 0)$.
\end{lemma}

Following Lemma \ref{lemma:update}, to solve (\ref{eq:update}), we must compute the eigendecomposition of
\begin{align*}
&\left( \sum_{i = 1}^n \sum_{l = 1}^n [\mathbf{D}_z]_{i, l}^2 \right)^{-1} \sum_{i = 1}^n \sum_{l = 1}^n [\mathbf{D}_z]_{i, l} \left(\Y_{i,:}^\top \Y_{l,:} - \sum_{k \neq z}[\mathbf{D}_k]_{i, l} \hat \bSigma_k^{(t)} \right) \\
&=\left( \underbrace{\sum_{i = 1}^n \sum_{l = 1}^n [\mathbf{D}_z]_{i, l}^2}_{\mathbf{Q}_{z, z}}  \right)^{-1} \left(\underbrace{\Y^\top \mathbf{D}_z \Y}_{\mathbf{W}_z} - \sum_{k \neq z} \hat \bSigma_k^{(t)} \underbrace{\sum_{i = 1}^n \sum_{l = 1}^n [\mathbf{D}_z]_{i, l} [\mathbf{D}_k]_{i, l}}_{\mathbf{Q}_{z, k}} \right).
\end{align*}
To expedite the computation, we can pre-compute $\mathbf{Q}_{k,k'}$ and $\mathbf{W}_{k}$ for all $k, k' = 1, \dots K$ as they do not depend on $\hat \bSigma_k^{(t)}$. If $q \gg n$, the eigendecomposition may be computationally prohibitive. However, the following proposition allows us to compute eigendecompositions of $n \times n$ matrices rather than $q \times q$ matrices in each iteration, simply by computing the compact SVD, $\Y = \mathbf{UDV}^\top$, up front, and computing the mvREHE estimator with data matrix $\Y\mathbf{V} \in \mathbb{R}^{n \times n}$. Thus, in practice, we must only compute eigendecompositions of $\min(n, q) \times \min(n, q)$ matrices.

\begin{prop}
Let $\Y = \mathbf{UDV}^\top$ be the compact SVD of $Y$. Then, the solution to \eqref{eq:mvREHEunpen} is $\mathbf{V}\tilde \bSigma_k \mathbf{V}^\top$ where $\tilde \bSigma_k$ is the solution to \eqref{eq:mvREHEunpen} using $\mathbf{YV}$ in place of $\Y$. 
\label{prop:runtime}
\end{prop}

Algorithm~\ref{alg:original} summarizes the procedure. The runtime of our algorithm is only $\mathcal{O}(nq^2 + n^2 \min(q, n)^2 + TK\min(q, n)^3)$, where $T$ is the number of iterations. In comparison, an \texttt{lme4}-based implementation of REML would involve operations scaling with at least $n^3 q^3$ in each iteration to simply evaluate the REML criterion. This is because REML involves computing the Cholesky decomposition of an $nq \times nq$ matrix \parencite{bates2015fitting}. Even the optimized algorithm for REML proposed by \textcite{zhou2014efficient}, which only applies to models with 2 components, or $K = 1$ in \eqref{eq:genmodel}, has a runtime complexity of $\mathcal{O}(n^3 + T_1 n q^2 + T_2 n q^6)$, where $T_1$ is the number of expectation-maximization iterations and $T_2$ is the number of Newton-Raphson iterations. The computational speed-up of our method makes it very appealing for high-dimensional and large sample applications. 

\begin{algorithm}[b]\caption{Blockwise coordinate descent algorithm for solving (\ref{eq:mvREHEunpen})} \label{alg:original}
\setstretch{1.15}
\textcolor{white}{X} 1. For $z \in [K]$, compute $\mathbf W_z = \Y^\top \mathbf{D}_z \Y$  \\
\textcolor{white}{X} 2. Compute $\mathbf Q$ where $\mathbf Q_{z,k} = \sum_{i = 1}^n \sum_{l = 1}^n [\mathbf{D}_z]_{i, l} [\mathbf{D}_k]_{i, l}$ for $z, k \in [K]$\\
\textcolor{white}{X} 3. For $t = 1, \dots, T$ iterations (or until convergence): \\
\textcolor{white}{XX} 3.1 For $z=1,\ldots,K$: \\
\textcolor{white}{XXX} 3.1.1 Compute eigendecomposition $\mathbf U \mathbf \Lambda \mathbf U^\top = (\mathbf W_z - \sum_{k \neq z} \mathbf Q_{z,k} \hat \bSigma_k^{(t)})/\mathbf Q_{z,z}$ \\
\textcolor{white}{XXX} 3.1.2 Update $\bSigma_z^{(t+1)} = \mathbf U \mathbf \Lambda^+ \mathbf U^\top$
\end{algorithm}


Since the unrestricted mvHE estimator in (\ref{eq:mvHE}) is obtained from a generalized estimating equation for observations $\Y_{i,j}\Y_{l,m}$ ($i,l \in [n], j,m \in [q]$) with an identity working correlation matrix, we can leverage results on generalized estimating equations \parencite{xie2003asymptotics} to establish asymptotic properties of our estimators. Specifically, under mild assumptions on the row covariance matrices $\mathbf{D}_k$, the mvHE estimator is consistent. Additionally, due to this consistency, if the true $\bSigma_k$ are positive definite, the mvHE and mvREHE estimators are asymptotically equivalent and are thus both consistent. We formally state these results in the following proposition, relegating the proof to Appendix \ref{sec:propproof}.
\begin{prop}
Suppose the model in \eqref{eq:genmodel} holds and the row covariance matrices $\mathbf{D}_k$ each satisfy $[\mathbf{D}_k]_{i, l} \leq \rho_h^*, h = |i - l|$, for a sequence $\{\rho_h^*\}_{h = 0}^\infty$ satisfying $\lim_{h \to \infty} \rho_h^* = 0$, and the true column covariance matrices are positive definite: $\bSigma_k \succ 0$. Then the mvHE and mvREHE estimators are consistent. \label{prop:consistency}
\end{prop}
The condition of Proposition~\ref{prop:consistency} is commonly satisfied in genetic studies, where the dataset consists of individuals from many unrelated families, i.e., the genetic relatedness ($\K_n$) and shared household ($\Hn_n$) matrices are approximately block diagonal up to permutation of the rows/columns. Despite the asymptotic equivalence of the mvHE and mvREHE estimators, our simulation results in Section \ref{sec:simulations} show a clear finite sample advantage to using mvREHE.

\subsection{Estimation of latent genetic regression parameters} \label{sec:regression_estimation}

Once we obtain an estimate $\hat{\bSigma}_G$, we can define plug-in estimates of the genetic regression parameters. However, rather than directly solving the plug-in version of \eqref{eq:geneticbeta_cov}, the high-dimensional nature of connectome data motivates the use of regularized estimators to improve prediction accuracy and prevent overfitting.

For each functional connection $j$, we consider the general framework where we solve:
\begin{equation} \label{eq:reg_regression}
\hat{\boldsymbol{\beta}}_{G,j} = \argmin_{\boldsymbol{\beta} \in C} \boldsymbol{\beta}^\top \hat{\bSigma}_G^{\mathcal{S},\mathcal{S}} \boldsymbol{\beta} - 2\boldsymbol{\beta}^\top \hat{\bSigma}_G^{\mathcal{S},\mathcal{F}_j} + P(\boldsymbol{\beta})
\end{equation}
where $P(\boldsymbol{\beta})$ is a penalty function and $C$ is a constraint set. This framework accommodates various regularization approaches, and we focus on three specific variants that are well-suited to connectome data.

The first two variants employ standard ridge and lasso penalties: $P(\boldsymbol{\beta}) = \lambda_1 \|\boldsymbol{\beta}\|_2^2$ for ridge and $P(\boldsymbol{\beta}) = \lambda_2 \|\boldsymbol{\beta}\|_1$ for lasso. These approaches address overfitting through shrinkage but do not leverage the matrix structure of brain connectivity. The ridge variant admits a closed-form solution, while the lasso variant can be efficiently solved using standard coordinate descent with active set and warm start optimizations.

The third variant, based on tensor regression, exploits the natural matrix organization of structural connectomes, where connections are arranged by their anatomical origin and destination regions. Rather than treating structural connections as an unstructured vector, this approach leverages their inherent spatial organization.

Specifically, let $\text{mat}(\mathbf{g}_i^{\mathcal{S}}) \in \mathbb{R}^{p \times p}$ denote the matrix representation of the genetic component of the structural connectome, where $p$ is the number of brain regions. For each functional connection $j$, we parameterize the linear predictor as $\langle \boldsymbol{\beta}_{1,j} \boldsymbol{\beta}_{2,j}^\top, \text{mat}(\mathbf{g}_i^{\mathcal{S}}) \rangle_F$, where $\boldsymbol{\beta}_{1,j}, \boldsymbol{\beta}_{2,j} \in \mathbb{R}^{p \times r}$ and $r$ is the rank parameter. This factorization allows connections that share anatomical endpoints to have related regression coefficients, reflecting the spatial organization of brain networks.

To achieve this structure, we define the constraint set as $C = \{\boldsymbol{\beta} : \text{mat}(\boldsymbol{\beta}) = \boldsymbol{\beta}_1 \boldsymbol{\beta}_2^\top, \boldsymbol{\beta}_1, \boldsymbol{\beta}_2 \in \mathbb{R}^{p \times r}\}$, which restricts the matrix form of $\boldsymbol{\beta}$ to have a low-rank factorization. In practice, we optimize over the factors $\boldsymbol{\beta}_{1,j}$ and $\boldsymbol{\beta}_{2,j}$ directly and add penalties on each factor for identifiability and shrinkage:
\begin{equation}\label{eq:tensor_regression}
\hat{\boldsymbol{\beta}}_{1,j}, \hat{\boldsymbol{\beta}}_{2,j} = \argmin_{\boldsymbol{\beta}_1, \boldsymbol{\beta}_2 \in \mathbb{R}^{p \times r}} \vect(\boldsymbol{\beta}_1\boldsymbol{\beta}_2^\top)^\top \hat{\bSigma}_G^{\mathcal{S},\mathcal{S}} \vect(\boldsymbol{\beta}_1\boldsymbol{\beta}_2^\top) - 2\vect(\boldsymbol{\beta}_1\boldsymbol{\beta}_2^\top)^\top \hat{\bSigma}_G^{\mathcal{S},\mathcal{F}_j} + \lambda\left(\|\boldsymbol{\beta}_1\|_F^2 + \|\boldsymbol{\beta}_2\|_F^2\right)
\end{equation}

We solve this optimization problem using alternating minimization: we  update $\boldsymbol{\beta}_{1}$ while holding $\boldsymbol{\beta}_{2}$ fixed, then update $\boldsymbol{\beta}_{2}$ while holding $\boldsymbol{\beta}_{1}$ fixed, and repeat until convergence. Each individual update admits a closed-form solution, making the optimization computationally efficient despite the non-convex nature of the overall problem. We then set $\hat{\boldsymbol{\beta}}_{G,j} = \vect(\hat{\boldsymbol{\beta}}_{1,j}\hat{\boldsymbol{\beta}}_{2,j}^\top)$.

Although we focused on latent genetic regression parameters, these parameters can similarly be defined for the common environmental and unique environmental components.
Moreover, while we focus on linear regression analysis here, the estimate $\hat{\boldsymbol{\Sigma}}_G$ serves as a summary statistic that can be used for many other multivariate data analyses of the genetic component $\mathbf{g}_i$ that are functions of estimated covariance matrices, including, for instance, principal component analysis and canonical correlation analysis.

\vspace{-3.5mm}

\section{Disentangling genetic and environmental contributions in the human connectome} \label{sec:application}
\vspace{-3.5mm}
\subsection{Data and family structure}

We analyze MRI data from 955 participants in the S1200 HCP dataset, including anatomical, diffusion-weighted, and resting-state functional MRIs preprocessed using the minimal preprocessing HCP pipeline \parencite{glasser2013minimal}. Cortical surfaces are registered across subjects using MSM surface matching \parencite{robinson2014msm}, then parcellated into 68 regions of interest (ROIs) using the Desikan-Killiany atlas \parencite{desikan2006automated}.

Our sample includes participants with varying genetic relatedness: monozygotic twins, dizygotic twins,  full siblings, half-siblings, and unrelated subjects. We encode this family structure using a $955 \times 955$ kinship matrix $\K_n$ representing the expected sharing of the genome, calculated from the HCP pedigree data using SOLAR-ECLIPSE \parencite{koran2014study} with kinship values of 1 for monozygotic twins, 0.5 for dizygotic twins and full siblings, 0.25 for half-siblings, and 0 for unrelated subjects. The shared environment matrix $\mathbf{H}_n$ is defined as $[\mathbf{H}_n]_{i, l} = \mathbbm{1}([\mathbf{K}_n]_{i, l} > 0)$, assuming siblings share the same household.

\vspace{-3.5mm}
\subsection{Connectome construction}

We used structural connectomes that were constructed from diffusion and structural MRI, as previously described \parencite{kiar2017highthroughput}. Using the minimally preprocessed HCP data, diffusion tensors were estimated in each voxel and streamline estimates of brain connection trajectories were propagated using the Euler delta crossings algorithm, as implemented in DiPy \parencite{garyfallidis2014dipy}. Each subject's anatomy was registered to the MNI152 atlas and fiber streamlines connecting each pair of regions in the Desikan-Killiany atlas were counted, resulting in a $68 \times 68$ structural connectome matrix for each subject. The subject-level connectomes used here are publicly available in the Neurodata portal \parencite{vogelstein2018communitydeveloped} and through the AWS Open Data program.

For functional connectomes, the HCP minimally preprocessed fMRI data provide a time-series at each location of the mid-thickness surface describing local neuronal activity. For each ROI, we compute a spatially averaged time-series describing the average neuronal activity pattern within that ROI using fMRI data from the first session. We use the resulting time-series to compute a $68 \times 68$ correlation matrix between ROIs, which we refer to as the functional connectome.

\vspace{-3.5mm}

\subsection{Statistical analysis}

\subsubsection{Estimation of covariance matrices} \label{subsec:estimationapplication}

While numerous representation models for functional and structural connectomes have been proposed in the literature \parencite{arsigny2006logeuclideana, arsigny2006geometric, dryden2009noneuclidean, zhang2019tensor, lila2022functional, zhao2022covarianceoncovariance}, we adopt a simple vectorization approach commonly used in neuroscience studies for its interpretability \parencite[see, for example,][]{smith2015linking,xia2018linked}. We represent each functional and structural connectome by vectorizing the full $68 \times 68$ connectivity matrices, yielding a 9,248-dimensional vector per subject (4,624 connections from each structural and functional matrix). Although this representation introduces redundancy due to symmetry in the connectivity matrices, it facilitates the tensor regression analysis without incurring a computational penalty in the covariance estimation step, as the covariance estimation runtime is dominated by the sample size rather than the feature dimension (see Proposition \ref{prop:runtime}). We use linear regression to adjust for the effects of age, age squared, sex, intracranial volume, and cortical volume, as in \textcite{colclough2017heritability}, and use the residuals as our mean-zero $\mathbf y_i$. While we do not enforce the matrix structure of connectomes at the stage of covariance estimation, we leverage this structure in the downstream latent genetic tensor regression analysis described in Section~\ref{sec:regression_estimation}.

We estimate the covariance matrices $\bSigma_k$ using our proposed mvREHE estimator, since multivariate REML is computationally intractable with $q = 9,248$ dimensions. Before applying mvREHE, we rescale the columns of $\Y$ to have unit standard deviation, then rescale the final estimates back to the original scale. This is equivalent to reweighting the loss function \eqref{eq:mvREHEunpen}, ensuring that estimation focuses on the correlation structure rather than being dominated by connections with larger raw variances. The reweighted estimator remains consistent for the true parameter.

\vspace{-3.5mm}

\subsubsection{Latent genetic and environmental regression analysis of structure-function coupling}

\textcite{gu2021heritability} previously investigated the genetics of structure-function coupling by first estimating the correlation between structural and functional connections for each subject---finding a moderate association with values up to 0.4--and then performing univariate heritability analysis on these resulting coefficients. That is, they estimate how much of the observed within-subject structure-function coupling can be attributed to genetic factors.

This moderate within-subject coupling does not translate to the population level, where the structural connectome shows virtually no linear predictive power for functional connections across subjects (Section \ref{subsec:datanalaysisresults}). In contrast, we hypothesize that a fundamental, genetically-encoded association between structure and function exists at the population level, but that it is obscured by unique environmental exposures.

Therefore, rather than asking, ``\textit{how heritable is an individual's summary of structure-function coupling?}'', as in \textcite{gu2021heritability}, we ask, ``\textit{what is the population-level linear relationship between structural and functional connectomes that is encoded in genetics?}'' By separating the genetic component from the environmental component, our methodology aims to uncover the structure-function coupling that can be inherited. This population-level framework also allows us to model more complex connectivity patterns in which a functional connection is influenced by multiple structural connections, which heritability analyses of subject-specific structure–function summaries cannot capture.

To study standardized effects, we apply the latent genetic regression framework described in Section~\ref{sec:regression_estimation}, but estimate the genetic regression parameters in \eqref{eq:geneticbeta} defined with $\tilde{\mathbf{g}}_i = \diag(\bSigma_G)^{-1/2} \mathbf{g}_i$ instead of $\mathbf{g}_i$. In practice, this means using the estimated genetic correlation matrix rather than the estimated genetic covariance matrix in the regularized loss function. This yields regression coefficients that represent the number of standard deviations the genetic component of functional connectivity changes for each standard deviation change in the genetic component of structural connectivity, holding the genetic component of all other structural connections fixed.

In our application to HCP data, we consider all three regularization methods (ridge, lasso, and tensor regression). To select the optimal regularization parameter and tuning parameters, we estimate the genetic correlation matrix and regression parameters on one half of the data and compute the genetic-$R^2$ on the other half, averaging the estimates of out-of-sample genetic-$R^2$ across splits, then choose the regularization method and tuning parameters that maximize this averaged out-of-sample genetic $R^2$. This approach allows us to determine which regularization strategy best captures the genetic structure-function relationships for each connection while avoiding overfitting.

We perform the same regression analysis on the common environment and unique environment components to compare linear predictability across latent components. This enables us to determine whether structure-function couplings are primarily driven by genetic factors, environmental factors, or a combination of both. We also perform classical linear regression using the observed data to compare against traditional analyses of connectome data and to assess how much the genetic decomposition reveals about structure-function coupling that would otherwise be masked in the observed data.

We report the out-of-sample genetic $R^2$ values for the selected method and tuning parameters. The resulting genetic $R^2$ values quantify the proportion of variance in each functional connection's genetic component that can be explained by a linear function of the genetic components of all structural connections. High genetic $R^2$ values indicate strong genetically-encoded structure-function coupling, while low values suggest little evidence of a linear relationship between the genetic components of structural and functional connectivity.

\subsection{Results} \label{subsec:datanalaysisresults}

\subsubsection{Genetic and environmental regression analysis of the structure-function coupling}

\begin{figure}
\centering
\includegraphics[width=\textwidth]{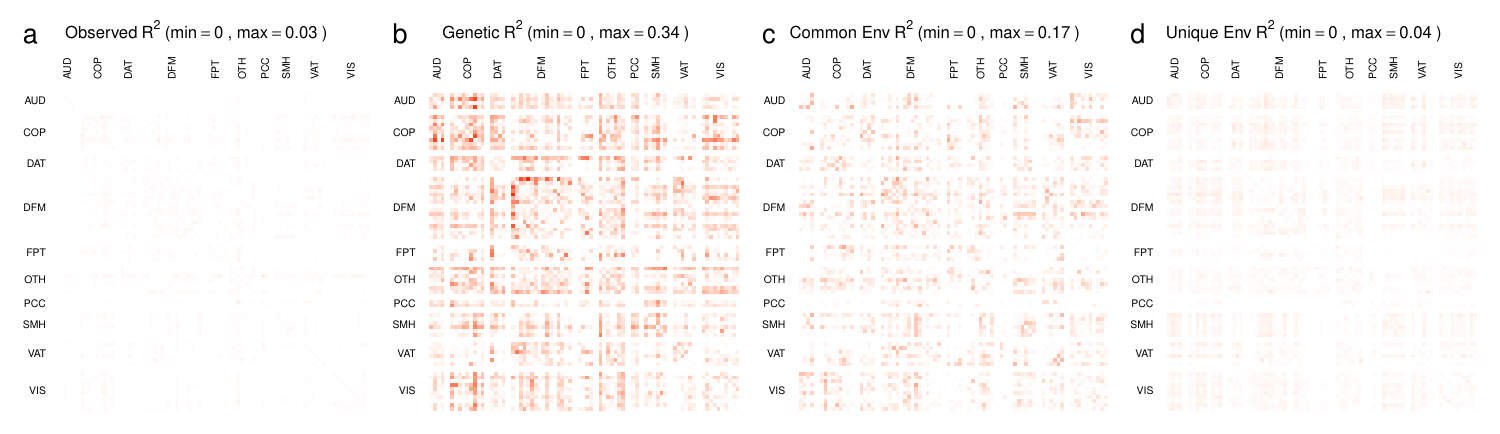}
\caption{Proportion of variance explained ($R^2$) in functional connectivity by structural connectivity for each functional connection. Results are shown for (a) observed data and decomposed latent components: (b) genetic, (c) common environment, and (d) unique environment. Each cell represents the out-of-sample $R^2$ for predicting one functional connection using all structural connections as predictors. Darker shades of red indicate higher predictability.}\label{fig:r2}
\end{figure}

Figure~\ref{fig:r2} shows the out-of-sample estimates of the proportion of variance of each functional connection that can be explained by all structural connections for the observed data (observed $R^2$) and each latent component (genetic $R^2$, common environment $R^2$, and unique environment $R^2$). Each $R^2$ value represents the best performance across all three regularized latent regression methods and their respective tuning parameters for that functional connection. The results reveal striking differences in structure-function predictability across observed and latent components.

In the observed data, structural connections show minimal linear predictive power for functional connections, with most $R^2$ values near zero and a maximum of only 0.03. However, when restricted to latent components, a clear pattern emerges. The genetic component shows the strongest structure-function coupling, with many functional connections exhibiting substantial predictability from structural connections (maximum $R^2$ = 0.34). The common environment component shows moderate predictability (maximum $R^2$ = 0.17), while the unique environment component shows virtually no linear relationship between structural and functional connectivity (maximum $R^2$ = 0.04).

These results suggest that the apparent lack of structure-function coupling in observed data masks a strong genetically-encoded relationship that is obscured by unique environmental factors affecting each individual differently.

Different mechanisms could explain the observed phenomenon of high predictability within the genetic component \parencite{vanrheenen2019genetic}, such as horizontal pleiotropy, where a shared set of genes contributes directly or through intermediate endophenotypes to both structural and functional connectivity. Alternatively, a causal relationship could exist, i.e., vertical pleiotropy, although the observed high predictability may also be due to spurious pleiotropy, such as biases in the imaging pipelines used to produce surrogate estimates of functional and structural connectivity.

\begin{figure}
\centering
\includegraphics[width=\textwidth]{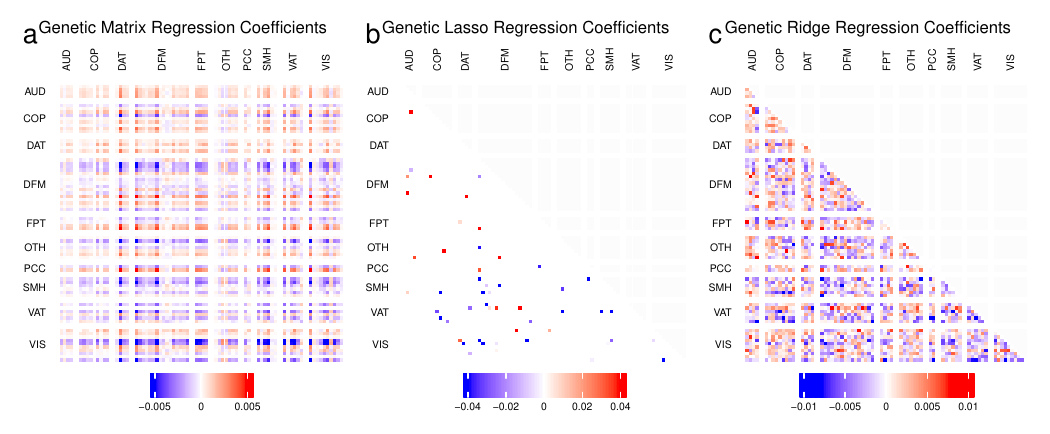}
\caption{Estimated genetic regression coefficients for predicting a functional connection within the default mode network (DFM) using all structural connections as predictors. Results shown for (a) tensor regression, (b) lasso regression, and (c) ridge regression. For ridge and lasso, coefficients from symmetric structural connection pairs are summed and displayed in the lower triangle. For tensor regression, the full coefficient matrix structure is preserved to illustrate the low-rank structure.}\label{fig:coefs}
\end{figure}

Figure~\ref{fig:coefs} shows the estimated genetic regression coefficients from all three regularization methods for a functional connection within the default mode network (DFM), which has the highest genetic $R^2$ (0.34). The three methods achieved different out-of-sample genetic $R^2$ values: ridge regression ($R^2$ = 0.34), lasso regression ($R^2$ = 0.27), and tensor regression ($R^2$ = 0.26). We display results from all three methods to contrast their regularization effects.

Since the structural connectome is symmetric, we handle the coefficient visualization differently across methods for interpretability. For ridge and lasso (panels a-b), we sum coefficients from symmetric pairs of structural connections since only their sum affects prediction, displaying results in the lower triangle with upper triangle set to zero. For tensor regression (panel c), we preserve the estimated coefficient matrix to illustrate the low-rank structure.

The coefficient patterns reflect each method's regularization properties: ridge produces distributed shrinkage across connections, lasso yields sparse selection, and tensor regression captures structured spatial relationships through its rank constraint. Interestingly, ridge regression achieves the highest performance for this DFM connection despite tensor regression's incorporation of the structural connectome's matrix structure, suggesting that distributed shrinkage may be more beneficial than structural constraints for this particular connection. 

Across all functional connections, tensor regression is optimal for 44.8\% of connections, ridge regression for 44.2\%, and lasso regression for 11.0\%, indicating that no single regularization approach dominates and that the optimal method depends on the specific functional connection being modeled. These differences demonstrate how regularization choice affects both the performance and interpretability of estimated structure-function coupling in the genetic component.


\vspace{-3.5mm}

\section{Simulations} \label{sec:simulations}

In this section, we consider a simulation study based on a related analysis of the HCP data, in an effort to validate our choice of methodology. That is, we simulate data from the additive genetics and common environment model in \eqref{eq:multivariatevariateobservedmodel}, where the row-covariance and column-covariance matrices are set based on estimates from a coarser-resolution analysis of the HCP data.

We group the 68 regions into 10 functional communities and analyze the connectivity at this resolution. For structural connectomes, we sum all structural connections within each community to obtain inter-community connection strengths. For functional connectomes, we average the time series of all regions within each community and compute correlations on these averaged time series. We then apply the mvREHE estimator to estimate the genetic, common environment, and unique environment covariance matrices, which serve as the true parameters for our simulation model.

The genetic relatedness matrix $\K_n$ is defined as follows: if the sample size $n \leq 1,000$, $\K_n$ is the top left $n \times n$ block of the HCP kinship matrix; if $n > 1,000$, then $\K_n$ is an $n \times n$ block diagonal matrix with blocks defined by the $1,000 \times 1,000$ HCP kinship matrix. As in Section \ref{subsec:estimationapplication}, we set the common household matrix $\Hn_n$ to be a binarized version of $\K_n$.

In each replicate, we sample observed data from the model in \eqref{eq:multivariatevariateobservedmodel} and estimate $\bSigma_k$ using mvHE and mvREHE. We do not use multivariate REML as it is computationally intractable with $q = 200$ dimensions. For comparison with existing univariate approaches, we also estimate the diagonal elements of $\bSigma_k$ using univariate HE, REHE, and REML estimators applied separately to each trait. We perform 50 replications for each simulation setting.

\begin{figure}[!t]
\centering
\includegraphics[width=\textwidth]{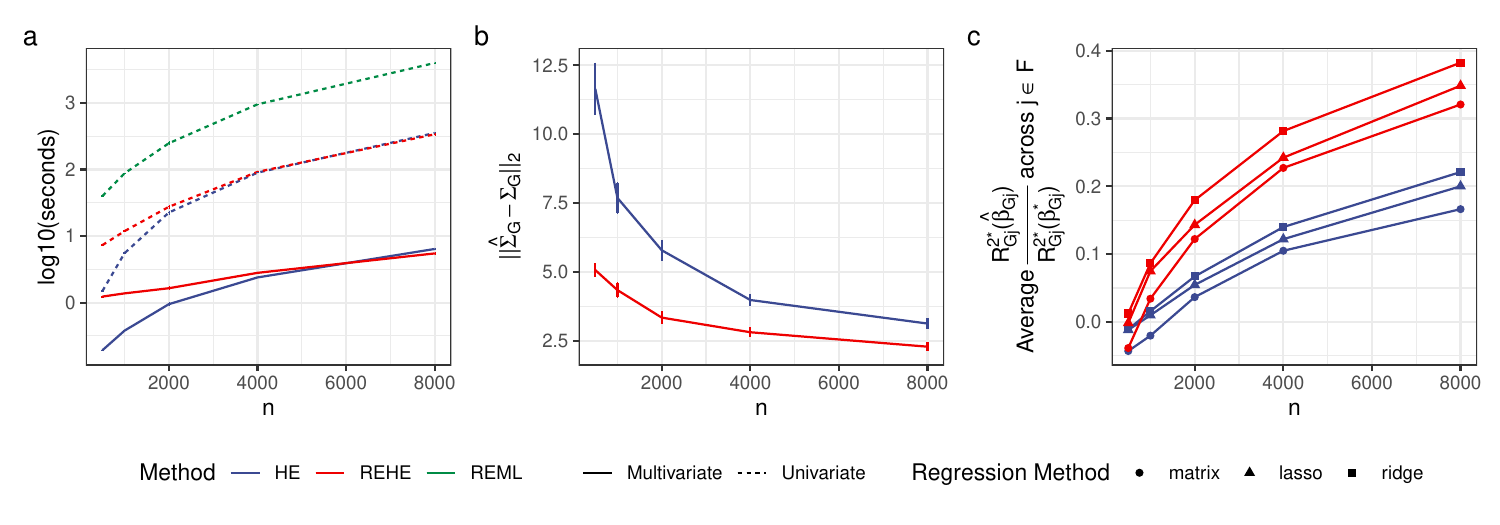}
\caption{Performance of the proposed estimation framework across different analyses. (a) Runtime comparison of multivariate methods (mvHE, mvREHE) versus univariate methods (HE, REHE, REML), (b) spectral-norm error of estimated genetic covariance matrix $\hat{\bSigma}_G$ comparing mvHE and mvREHE, and (c) prediction performance of latent genetic regression analysis showing the average ratio of achieved genetic-$R^2$ to oracle genetic-$R^2$ across functional connections for different combinations of covariance estimation methods and regularization approaches, all with varying sample size $n$ and $q = 200$ fixed.}\label{fig:simdatacombined}
\end{figure}

In Figure~\ref{fig:simdatacombined}a, we compare the runtime of multivariate and univariate estimators. The multivariate methods mvHE and mvREHE show virtually identical computational times, with execution times under a couple of seconds even at $n = 8,000$. The positive definite constraint in mvREHE adds no computational burden due to our efficient algorithm. Both multivariate methods are orders of magnitude faster than multiple univariate estimators, particularly REML, while providing full covariance matrices rather than just diagonal elements.

We next compare the estimation error of the entire covariance matrix using mvHE and mvREHE, the only computationally feasible multivariate estimators in this setting. Figure~\ref{fig:simdatacombined}b shows results for the genetic component, with results for other components in Figure~\ref{fig:simdataspectral} in the Supplementary Materials. The positive semi-definite constraint in mvREHE yields considerably lower estimation error than the truncation-based mvHE, particularly at smaller sample sizes, while having nearly identical runtime performance.

Finally, we compare the prediction performance of latent genetic regression analysis using mvHE and mvREHE for the plug-in covariance estimates. We evaluate all combinations of covariance estimation methods (mvHE vs mvREHE) and regularization approaches (ridge, lasso, tensor regression). Figure~\ref{fig:simdatacombined}c shows the average ratio of achieved genetic-$R^2$ to the oracle genetic-$R^2$ across all functional connections, where the oracle represents the theoretical maximum achievable with the true covariance matrix. Values closer to 1 indicate better recovery of the true predictive relationships in the genetic component. The results demonstrate that mvREHE consistently outperforms mvHE across all regularization methods, with the performance gap particularly pronounced at  larger sample sizes. Results for the common environmental and unique environmental components are shown in Figure~\ref{fig:simdatabeta}. This confirms that the improved covariance estimation accuracy of mvREHE translates directly into better downstream prediction performance across all latent components.

As it was not computationally feasible to directly compare mvREHE with the multivariate REML estimator with $q = 200$, we perform additional low-dimensional simulations $(q \leq 10)$ in Section \ref{subsec:lowdim} of the Supplementary Materials. In Figure \ref{fig:sim1combined}, we observe that the estimation performance of mvREHE nearly exactly matches that of multivariate REML, both for estimation of the entire covariance matrix as well as heritability proportions, while being nearly three orders of magnitude faster than the when $q = 5$ and five orders of magnitude faster when $q = 10$. 

In summary, these simulation results suggest that the mvREHE estimator provides the best of both worlds in terms of computational efficiency and estimation accuracy in the data-inspired settings considered.

\vspace{-3.5mm}

\section{Discussion}\label{sec:discussion}

We developed a novel latent genetic regression analysis framework that characterizes relationships between genetic components of brain connectivity. Unlike previous approaches that estimate subject-specific structure-function coupling and then assess its heritability \parencite{gu2021heritability}, our method directly models genetic contributions, additionally allowing for a population-level understanding of complex interaction patterns where functional connections can be influenced by multiple structural connections.

The computational feasibility of this analysis required methodological innovation in multivariate variance component estimation. Existing estimators like multivariate REML are computationally intractable for high-dimensional connectome data. We introduce a constrained method-of-moments estimator that guarantees positive semi-definite covariance estimates. Simulation studies demonstrate that our block coordinate descent algorithm efficiently incorporates the positive semi-definite constraint while maintaining practically identical runtime as the unconstrained estimator, and that the method nearly matches multivariate REML's estimation accuracy while providing dramatic computational speedups.

Our analysis reveals that functional connectivity is substantially more predictable from structural connectivity when restricted to genetic components (maximum $R^2$ = 0.34) compared to observed data, which show minimal predictive relationships (maximum $R^2$ = 0.03). This finding suggests that unique environmental factors mask genetically-encoded relationships between structural and functional connectomes.

The proposed method extends well beyond neuroscience applications, applying to any multivariate setting where observed data represents the sum of latent genetic and environmental components. This enables the analysis of complex, high-dimensional phenotypes which can be modelled as the sum of independent latent components, which was previously computationally infeasible. By estimating entire covariance matrices rather than just variances, our method enables a diverse set of multivariate analyses restricted to individual latent components, including principal component analysis, canonical correlation analysis, and the regression framework demonstrated here. These analyses fundamentally rely on covariance structure.

\section{Code availability}

An open-source R package implementing the proposed estimator is available at 
\url{https://github.com/keshav-motwani/mvREHE}.


\printbibliography

\newpage 
\spacingset{1}

\appendix
\appendixpage

\section{Proof of Lemma 1} \label{prooflemma}

\begin{proof}
Expressing the Frobenius norm as the trace of a quadratic form, using linearity of the trace operator, and applying an eigendecomposition gives us that
\begin{align*}
\hat \bSigma &= \argmin_{\bSigma \succeq 0}  \sum_{w = 1}^W \|\mathbf{S}_w - a_w \bSigma \|_F^2 \\
&= \argmin_{\bSigma \succeq 0}   \left\|\left(\sum_{w = 1}^W a_w^2\right)^{-1} \sum_{w = 1}^W a_w \mathbf{S}_w - \bSigma \right\|_F^2\\
&=\argmin_{\bSigma \succeq 0}   \left\|\mathbf S - \bSigma \right\|_F^2 \\
&=\argmin_{\bSigma \succeq 0}   \left\|\mathbf U \mathbf \Lambda \mathbf U^\top  - \bSigma \right\|_F^2 \\
&=\argmin_{\bSigma \succeq 0}   \left\|\mathbf \Lambda  - \mathbf U ^\top \bSigma \mathbf U \right\|_F^2.
\end{align*}
Since $\mathbf \Lambda$ is diagonal, we can restrict ourselves to the case where $\mathbf U ^\top \bSigma \mathbf U$ is also diagonal. Note that the optimal $\bSigma$ is PSD if and only if $\mathbf U ^\top \bSigma \mathbf U$ is PSD, since $\mathbf U$ is an orthogonal matrix. Thus, the diagonal elements of $\mathbf U ^\top \bSigma \mathbf U$ must be non-negative. If $\lambda_j \geq 0$, then the best case is that the $j$th diagonal element of $\mathbf U ^\top \bSigma \mathbf U$ is also $\lambda_j$. If $\lambda_j < 0$, then the best case is that the $j$th diagonal element of $\mathbf U ^\top \bSigma \mathbf U$ is $0$. Hence, the minimizer is achieved by $\hat \bSigma = \mathbf U \mathbf \Lambda^+ \mathbf U^\top$. 
\end{proof}

\newpage

\section{Proof of Proposition \ref{prop:consistency}} \label{sec:propproof}
\begin{proof}
First, we define some notation. Let $\dot Y = \vect(\Y^\top)$. Then under the model in equation (\ref{eq:genmodel}), $\dot Y \sim \mathcal{N}(0, \sum_{k = 0}^K \mathbf D_k \otimes \bSigma_k)$. Let $\mathbf E_{jl}$ is a $q$ by $q$ matrix with 1 in the $j,l$ and $l,j$ entries and 0 elsewhere. Then equation (\ref{eq:mvHE}) can be written as the solution to
\begin{align}\label{eq:rewrite}
&\argmin_{\bSigma_k} \| \dot \Y \dot \Y^\top - \sum_{k = 0}^K \mathbf D_k \otimes \bSigma_k \|_F^2 \\
&= \argmin_{\bSigma_k} \| \dot \Y \dot \Y^\top -  \sum_{k = 0}^K \sum_{j \geq l} [\bSigma_k]_{j, l} (\mathbf D_k \otimes \mathbf E_{jl}) \|_F^2 \\
&= \argmin_{\sigma} \| \tilde \Y - \tilde{\mathbf{X}} \boldsymbol \sigma \|_2^2
\end{align}
where $\boldsymbol \sigma_{g(k, j, l)} := [\bSigma_k]_{j,l}$, $\tilde \Y = \vect(\dot \Y \dot \Y^\top)$, and $\tilde{\mathbf{X}}_{:, g(k, j, l)} = \vect(\mathbf D_k \otimes \mathbf E_{jl})$ and $g$ is a function which linearizes the indices $k \in \{0\} \cup [K]$, $j, l \in [q]$ with $j \geq l$: $$g(k, j, l) = \frac{kq(q+1)}{2} + (l - 1)(q + 1) - \frac{l(l - 1)}{2} + j - (l - 1).$$ Therefore, (\ref{eq:rewrite}) is the ordinary least squares estimate of $\sigma$ in the linear model
$$
\tilde \Y = \tilde{\mathbf{X}} \boldsymbol \sigma + \boldsymbol \delta
$$
where $\cov(\boldsymbol \delta) = \mathbf R$. We can see that the pre-truncation mvHE estimator is the special case of a generalized estimating equation estimator with the identity matrix as the working correlation matrix. Thus, as in \textcite{yue2021rehe}, we use results from \textcite{xie2003asymptotics} on asymptotics of generalized estimating equations. 

One can verify that Assumption 1 implies that the correlation of $\boldsymbol \delta$ is bounded by $|\mathbf R_{a, b} / \sqrt{\mathbf R_{a, a} \mathbf R_{b, b}}| < \rho_h$ with $h = |a - b|$ for a sequence $\rho_h$ with $\lim_{h \to \infty} \rho_h = 0$, since 
\begin{align*}
\cov(\Y_{i, j}\Y_{l, m}, \Y_{i', j'}\Y_{l', m'}) &= E(\Y_{i, j}\Y_{l, m}\Y_{i', j'}\Y_{l', m'}) - E(\Y_{i, j}\Y_{l, m})E(\Y_{i', j'}\Y_{l', m'}) \\ 
&= E(\Y_{i, j}\Y_{i', j'})E(\Y_{l, m}\Y_{l', m'}) + E(\Y_{i, j}\Y_{l', m'}) E(\Y_{l, m}\Y_{i', j'}) \\
&= \cov(\Y_{i, j},\Y_{i', j'})\cov(\Y_{l, m},\Y_{l', m'}) + \cov(\Y_{i, j},\Y_{l', m'}) \cov(\Y_{l, m},\Y_{i', j'})
\end{align*}
and 
$$
\cov(\Y_{i, j}, \Y_{l, m}) = \mathbb{E}(\Y_{i, j} \Y_{l, m}) = [\bSigma_E]_{j, m} [\I_n]_{i, l} + [\bSigma_G]_{j, m} [\K_n]_{i, l}.
$$
Therefore, under this assumption, the pre-truncation mvHE estimator is consistent \parencite{xie2003asymptotics, yue2021rehe}. Additionally, the truncated mvHE estimator is consistent by the continuous mapping theorem.

Now we argue asymptotic equivalence of the mvHE and mvREHE estimators if the true $\bSigma_k$ are positive definite. Note that the mvREHE estimate is the same as the mvHE estimate if the covariance matrices estimated by mvHE are PSD. By openness of the space of positive definite matrices, there exists an open ball of positive definite matrices around the true $\bSigma_k$ which the mvHE estimate will asymptotically belong to, by consistency. Therefore, the mvREHE estimate will asymptotically be the same as the the mvHE estimate, and will also be consistent. 
\end{proof}

\renewcommand{\thefigure}{S\arabic{figure}}

\section{Supplemental Simulation Results} 
\label{sec:suppfigs}

\subsection{Data-inspired simulation (described in Section \ref{sec:simulations} of main text)}

\begin{figure}[!htb]
\centering
\includegraphics[width=\textwidth]{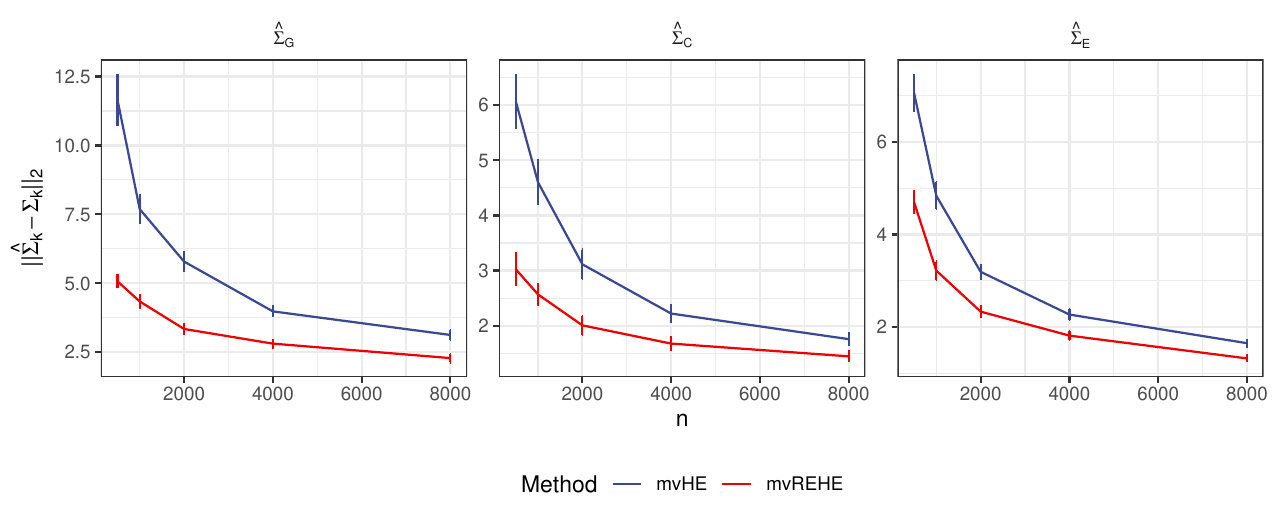}
\caption{Estimation error of all three covariance matrices $\bSigma_k$ (genetic, common environmental, and unique environmental components) comparing mvHE and mvREHE estimators with varying sample size $n$. Error is measured using the spectral norm of the difference between estimated and true covariance matrices. Results demonstrate the superior performance of mvREHE over mvHE across all components, particularly at smaller sample sizes.}
\label{fig:simdataspectral}
\end{figure}

\begin{figure}[!htb]
\centering
\includegraphics[width=\textwidth]{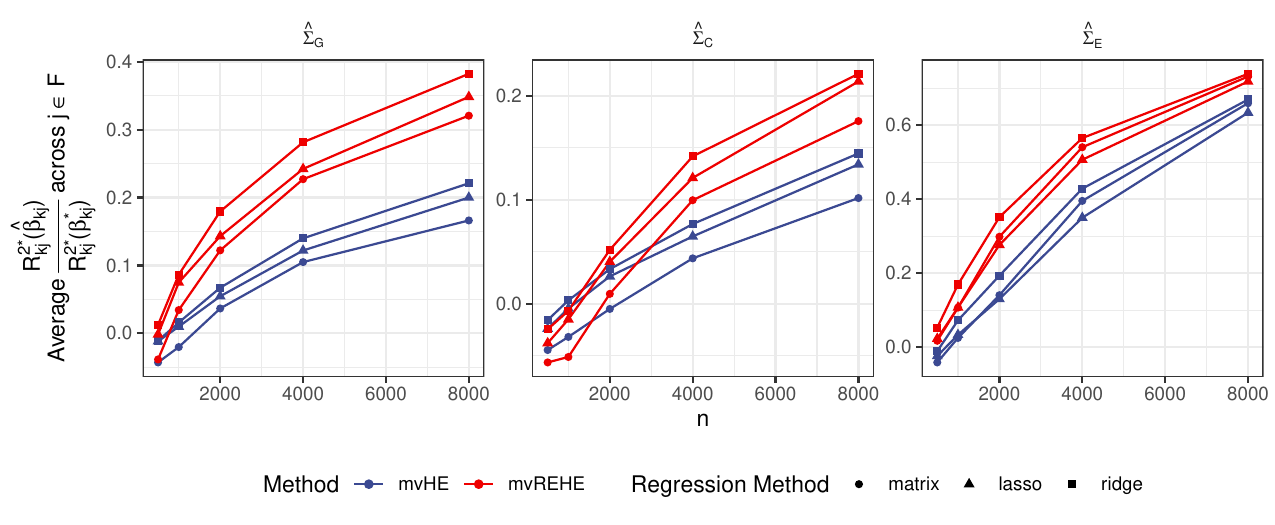}
\caption{Prediction performance of latent regression analysis for common environmental and unique environmental components. Results show the average ratio of achieved $R^2$ to oracle $R^2$ across all functional connections for different combinations of covariance estimation methods (mvHE vs mvREHE) and regularization approaches (ridge, lasso, tensor regression) with varying sample size $n$.}
\label{fig:simdatabeta}
\end{figure}


\subsection{Low-dimensional simulation} \label{subsec:lowdim}

In this section, we perform simulations in the low-dimensional setting, where it is possible to directly compare with the multivariate REML estimator. Due to its statistical efficiency, the multivariate REML estimator is commonly considered a gold standard, despite being computationally inefficient. We set the covariance matrices $\bSigma_k$ to be scaled unstructured correlation matrices with $q = 5, 10, 20$ traits. The scaling is defined such that the multi-trait heritability matches that of the estimated model from the real data application. The unstructured correlation matrices are fixed to be a uniform draw from the space of correlation matrices, across all simulation replicates. We consider sample sizes ranging from 250 to 8000.


\begin{figure}
\centering
\includegraphics[width=\textwidth]{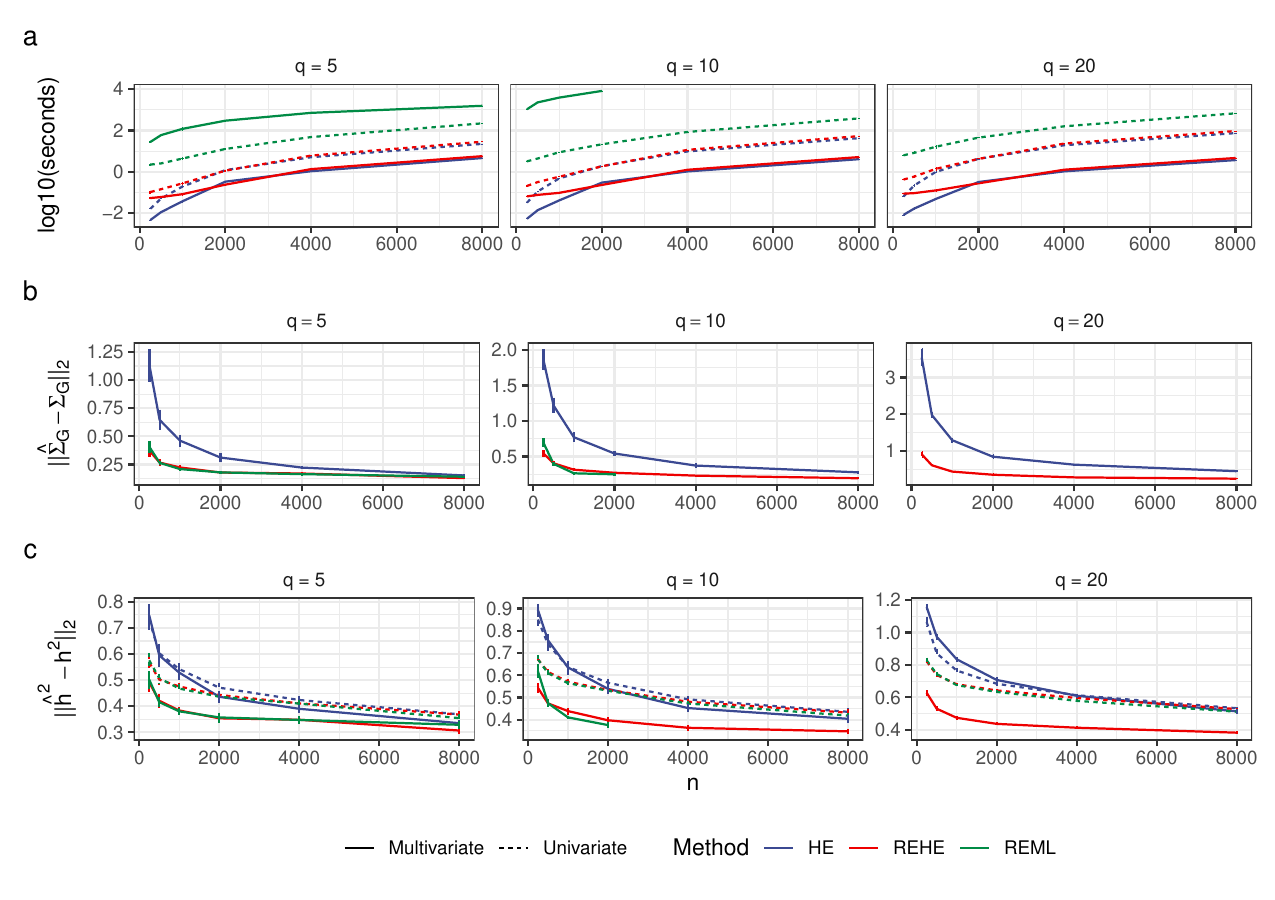}
\caption{(a) Runtime, (b) spectral-norm error of $\hat \bSigma_G$, and (c) Euclidian-norm error of the individual-trait heritability proportions $h^2$ in the setting of unstructured, low-dimensional $\bSigma_k$ with varying $q$ and $n$.}\label{fig:sim1combined}
\end{figure}

In Figure~\ref{fig:sim1combined}a, we show the runtime for each of the estimators considered. First, we can see that mvREHE has a similar runtime to mvHE --- that is, imposing the positive semi-definite constraint does not affect the runtime in these settings. We can also see that mvREHE is nearly three orders of magnitude faster than the multivariate version of REML when $q = 5$ and nearly five orders of magnitude faster when $q = 10$. With large sample sizes and $q = 10$ or $q = 20$ with any sample size, the multivariate REML estimator cannot be computed in reasonable time; thus, these results are not shown. We can see that increasing the number of traits $q$ does not have a large impact on the runtime of mvREHE, while for multivariate REML the increase in runtime is substantial. Additionally, the multivariate REML estimator is susceptible to convergence issues, resulting in a large number of replicates returning no estimates for this estimator (over 90\% of replicates failing in some settings with large sample sizes; see Figure~\ref{fig:mvremlfail} in the Appendix). Finally, the mvREHE estimator is even faster than fitting separate univariate models for each trait using any of the univariate estimators, while estimating the entire covariance matrix rather than the diagonal only.

Given that the mvREHE estimator is the most computationally efficient, one may ask if it compromises on estimation performance. In Figure~\ref{fig:sim1combined}b, we show the spectral error of estimating the covariance matrix of the genetic component, with results for the other components given in Figure~\ref{fig:sim1spectral}. We can see that the PSD constraint in the mvREHE estimator results in a large improvement over the mvHE estimator, nearly exactly matching the estimation accuracy of the multivariate REML estimator in all cases where the REML estimate could be computed. 

We also compare estimation performance of the individual-trait heritability proportions in Figure~\ref{fig:sim1combined}c, as measured by the Euclidian-norm error on the vector of heritability proportion of all traits. Since the heritability proportion depends only on the diagonal elements of the covariance matrices, we compare against the univariate estimators as well. We again see that mvREHE and multivariate REML have nearly identical performance, and that these estimators greatly outperform their commonly-used univariate counterparts. Finally, we see a large benefit from using a PSD constraint in mvREHE rather than simply truncating the estimate in mvHE. 


In Figure~\ref{fig:sim1combined2component}, we present the results of a simulation akin to the one described in this section, but with only the genetic and unique environment components, in order to facilitate comparison with the optimized REML algorithm proposed by \textcite{zhou2014efficient} which only applies to two component models. In the two component setting, all methods have very similar estimation performance. While the algorithm proposed by \textcite{zhou2014efficient} is up to two orders of magnitude faster than the \texttt{lme4}-based implementation of REML with $q = 10$, it is still two orders of magnitude slower than mvREHE when $q = 10$ and four orders of magnitude slower than mvREHE when $q = 20$.

\setcounter{figure}{0}

\begin{figure}[!htb]
\centering
\includegraphics[width=\textwidth]{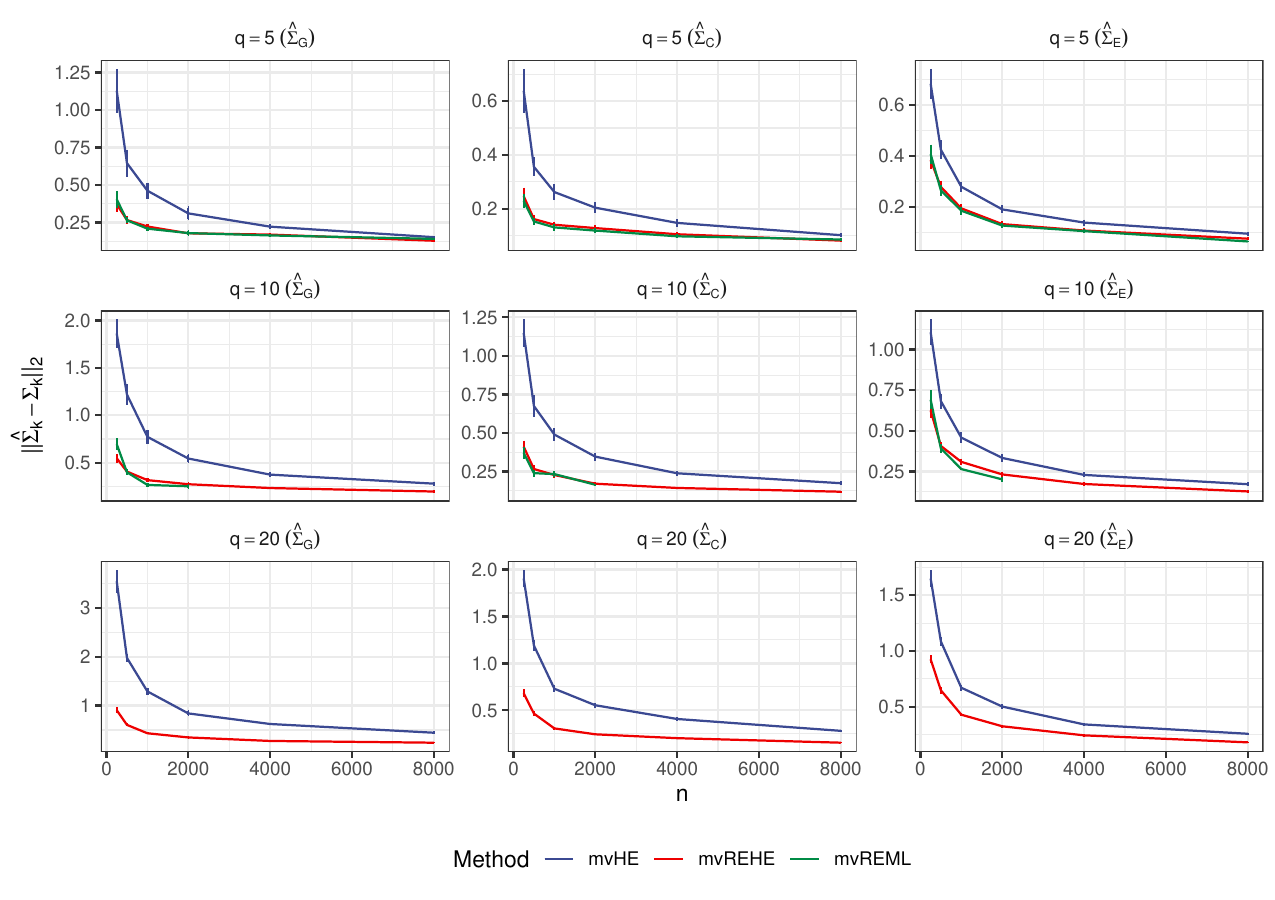}
\caption{Estimation error of all three covariance matrices $\bSigma_k$ in the setting of low-dimensional $\bSigma_k$
with varying sample size $n$.}
\label{fig:sim1spectral}
\end{figure}

\begin{figure}[h]
\centering
\includegraphics[width=0.8\textwidth]{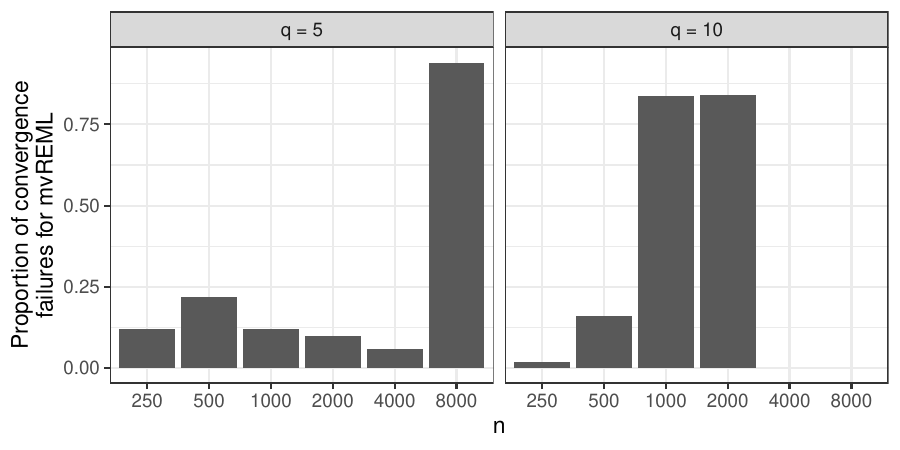}
\caption{Proportion of simulation replicates in low-dimensional simulation setting in which the multivariate REML estimator fails to converge. Note that no results are reported for the settings with $q=10$ and $n = 4000$ and $8000$, as REML estimation is computationally prohibitive in these cases.}\label{fig:mvremlfail}
\end{figure}


\begin{figure}[h]
\centering
\includegraphics[width=\textwidth]{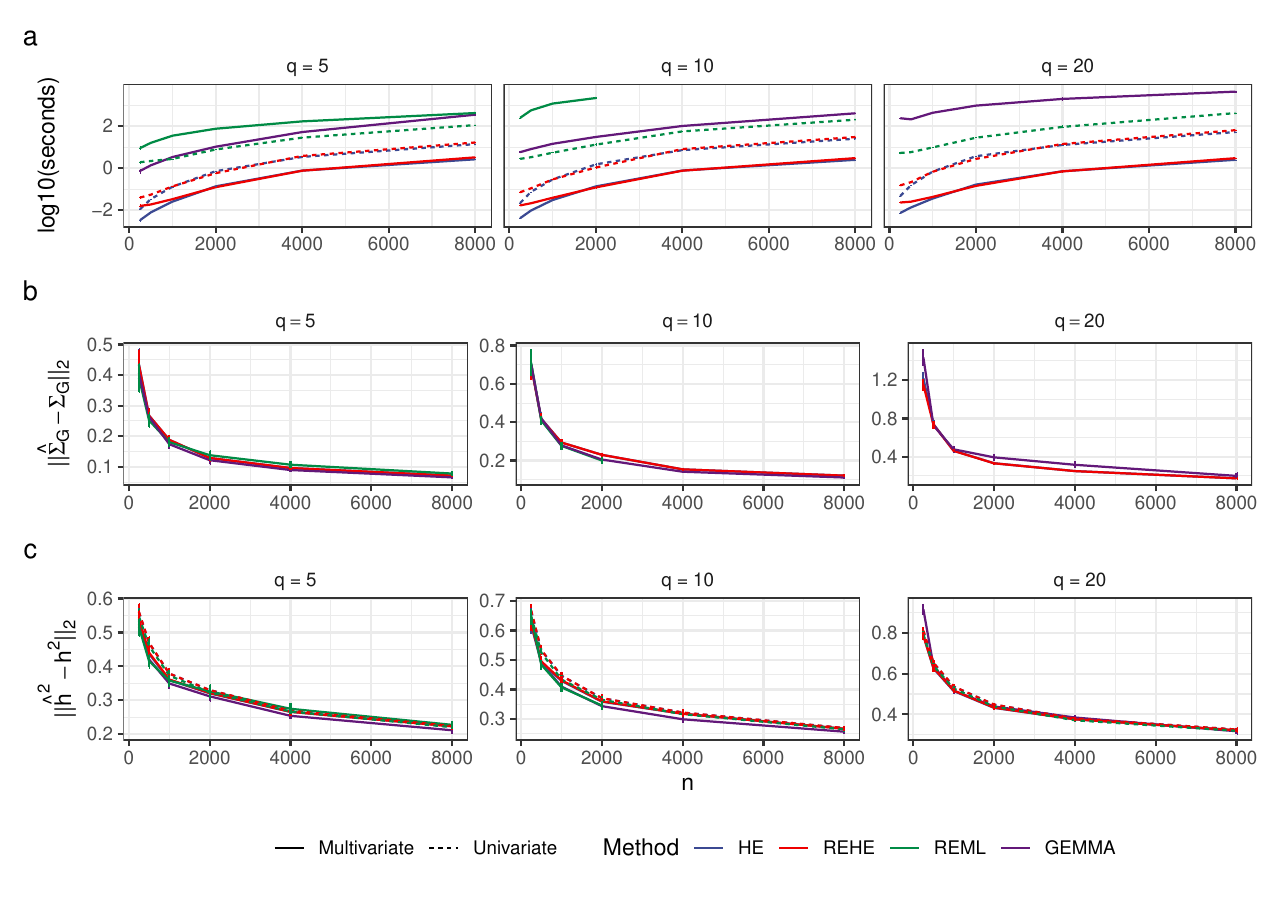}
\caption{(a) Runtime, (b) spectral-norm error of $\hat \bSigma_G$, and (c) Euclidian-norm error of the individual-trait heritability proportions $h^2$ in the setting of unstructured, low-dimensional $\bSigma_k$ with varying $q$ and $n$ with only the genetic and unique environmental components (two component model).  GEMMA is the optimized algorithm for computing the multivariate REML estimator, proposed by \textcite{zhou2014efficient}. }\label{fig:sim1combined2component}
\end{figure}


\section{Extension to function-valued traits} \label{sec:functional}

Often times, neuroimaging traits can be measured along a continuum, though only discrete measurements are obtained. Here, we show how the proposed mvREHE estimator can be easily combined with ideas from functional data analysis \parencite{yao2005functional} to incorporate smoothness into the covariance estimation step, similar to \textcite{risk2021ace}.

Consider the following model for the noiseless data of the $i$th subject:
$$
y_i(t) = g_i(t) + c_i(t) + e_i(t)
$$
for $t \in [0, 1]$, where $g_i$, $c_i$, and $e_i$ are mean-zero Gaussian processes with smooth covariance functions $C_G$, $C_C$, and $C_E$ respectively. Suppose we observe noisy versions of $y_i$ at $q$ discrete points $t_1, \dots, t_q$, denoted as the vector $\mathbf{y}_i$, where
\begin{equation} \label{eq:funcmodel}
[\mathbf{y}_i]_j = y_i(t_j) + \delta_{ij} = g_i(t_j) + c_i(t_j) + e_i(t_j) + \delta_{ij}
\end{equation} 
for $j = 1, \dots, q$, where $\delta_{ij} \sim \mathcal{N}(0, \sigma^2)$ are i.i.d. Then,
\begin{equation}  \label{eq:multivariateindivmodel}
\mathbf{y}_i \overset{d}{=} \mathbf{g}_i + \mathbf{c}_i + \mathbf{e}_i + \boldsymbol{\delta}_i
\end{equation}
where
$\mathbf{g}_i \sim \mathcal{N}(0, \bSigma_G)$, $\mathbf{c}_i \sim \mathcal{N}(0, \bSigma_C)$, $\mathbf{e}_i \sim \mathcal{N}(0, \bSigma_E)$, and $\boldsymbol{\delta}_i \sim \mathcal{N}(0, \sigma^2 \I_q)$, and $[\bSigma_k]_{j,l} = C_k(t_j, t_l)$ for $k \in \{G, C, E\}$. 

Let $\mathbf{Y}, \mathbf{G}, \mathbf{C}, \mathbf{E}$, and $\boldsymbol{\Delta}$ be matrices with rows $\mathbf{y}_i$,  $\mathbf{g}_i$, $\mathbf{c}_i$, $\mathbf{e}_i$, and $\boldsymbol{\delta}_i$, respectively. As in the observed data model in equation \eqref{eq:multivariatevariateobservedmodel}, we assume that $\G \sim \mathcal{MN}(\mathbf{0}_{n \times q}, \K_n, \bSigma_G)$, $\C \sim \mathcal{MN}(\mathbf{0}_{n \times q}, \Hn_n, \bSigma_C)$, $\E \sim \mathcal{MN}(\mathbf{0}_{n \times q}, \I_n, \bSigma_E)$, and $\boldsymbol{\Delta} \sim \mathcal{MN}(\mathbf{0}_{n \times q}, \I_n, \sigma^2 \I_q)$. Without any additional assumptions, with $\mathbf D_0 = \I_n$, it is not possible to identify $\E$ and $\boldsymbol{\Delta}$, since they both have the same row-covariance, therefore $\E + \boldsymbol{\Delta}$ also follows a matrix-normal distribution with the same row-covariance. However, in the functional data setting described, it is reasonable to assume $C_{k}$ are smooth covariance functions, thus it is possible to estimate them even in the presence of noise. Specifically, we propose to fit model \eqref{eq:multivariatevariateobservedmodel}, but, similar to \parencite{yao2005functional, risk2021ace}, apply a 2D local polynomial smoother to the estimates $\hat \bSigma_{k}$ with the diagonal removed, to obtain estimates of $C_{k}$.

\subsection{Simulation results}

In this section, we simulate data from the model in equation (\ref{eq:funcmodel}) with covariance functions defined by
$$
C_G(s, t) = C_E(s, t) = \sum_{k = 1}^{50} k^{-2\alpha}\cos(k \pi s)\cos(k \pi t).
$$
We consider $\alpha = 1$ and $\alpha = 2$, where the covariance function with $\alpha = 2$ is smoother than that with $\alpha = 1$. We set the variance of the observational noise $\sigma^2 = 1$. We sample the function at $q = 100$ timepoints equally spaced between $0$ and $1$. 

To estimate the covariance functions, we define the mvHE and mvREHE estimates as piecewise-constant functions given by the elements of the estimated $\hat \bSigma_k$. We compare these to smoothed versions of these estimators, as described in Section \ref{sec:functional}. To get smooth estimates $\hat C_k$, we apply a local linear smoother to the estimated $\hat \bSigma_k$ after removing the diagonal elements, with an Epanechnikov kernel and bandwidth chosen by generalized cross-validation. 

\begin{figure}
\centering
\includegraphics[width=\textwidth]{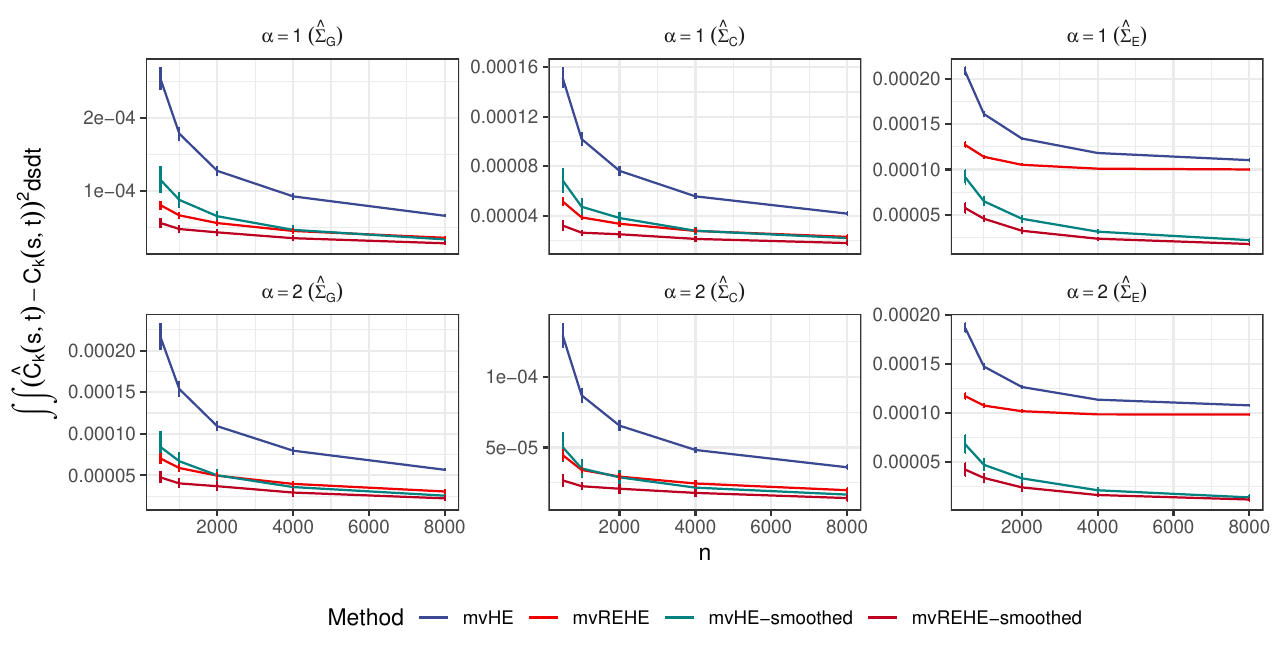}
\caption{Estimation error with two covariance functions with different degrees of smoothness, where larger $\alpha$ means smoother. The number of discrete observations along the function is $q = 100$ with $n$ varying.}\label{fig:sim3error}
\end{figure}

In Figure~\ref{fig:sim3error}, we show the integrated squared error of estimating $C_k$. We see that with low sample sizes, smoothing helps in estimation of the covariance functions for all components, and mvREHE improves over mvHE for both the smoothed and unsmoothed versions. For estimation of $C_E$, smoothing helps even with large sample sizes, since it is able to recover the smooth covariance function which does not contain the observational noise along the diagonal term, as described in Section \ref{sec:functional}. Note that the unsmoothed estimates of $C_E$ are not consistent due to the lack of identifiability without smoothness. This highlights the versatility of the proposed framework and its ability to be combined with other covariance estimation approaches to incorporate additional information about the data.

\end{document}